\documentclass[final,5p,times,twocolumn]{elsarticle}

\usepackage{amssymb}
\usepackage{url}

\usepackage{siunitx, amsmath, booktabs, bm}
\DeclareSIUnit\bar{bar}

\usepackage{pgfplots}
\usetikzlibrary{arrows.meta}
\pgfplotsset{compat=1.16}
\usepgfplotslibrary{groupplots}
\usepgfplotslibrary{colorbrewer}
\pgfplotsset{cycle list/Dark2}

\graphicspath{{figs/},}

\journal{Journal of Non-Newtonian Fluid Mechanics}

\begin{document}

\begin{frontmatter}



\title{Non-isothermal direct bundle simulation of SMC compression molding with a non-Newtonian compressible matrix\tnoteref{aam}}

\tnotetext[aam]{
NOTICE: this is the author’s version of a work that was accepted for publication in Journal of Non-Newtonian Fluid Mechanics. Changes resulting from the publishing process, such as peer review, editing, corrections, structural formatting, and other quality control mechanisms may not be reflected in this document. Changes may have been made to this work since it was submitted for publication. A definitive version was subsequently published in Journal of Non-Newtonian Fluid Mechanics, 310, 104940, (December 2022), \url{https://doi.org/10.1016/j.jnnfm.2022.104940}}


\author[FAST,UNIA]{Nils Meyer}
\author[FAST,ICT]{Sergej Ilinzeer}
\author[UWO]{Andrew N. Hrymak}
\author[FAST,ICT]{Frank Henning}
\author[FAST]{Luise Kärger}

\affiliation[FAST]{organization={Karlsruhe Institute of Technology (KIT), Institute of Vehicle System Technology},
            city={Karlsruhe},
            state={BW},
            country={Germany}}
            
\affiliation[UNIA]{organization={University of Augsburg, Institute of Materials Resource Management},
            city={Augsburg},
            state={BY},
            country={Germany}}

\affiliation[ICT]{organization={Fraunhofer Institute for Chemical Technology (ICT)},
            city={Pfinztal},
            state={BW},
            country={Germany}}
            
\affiliation[UWO]{organization={Western University, Dept. of Chemical \& Biochemical Engineering},
            city={London},
            state={ON},
            country={Canada}}

\begin{abstract}
Compression molding of Sheet Molding Compounds (SMC) is a manufacturing process in which a stack of discontinuous fiber-reinforced thermoset sheets is formed in a hot mold. 
The reorientation of fibers during this molding process can be either described by macroscale models based on Jeffery's equation or by direct mesoscale simulations of individual fiber bundles. 
In complex geometries and for long fibers, direct bundle simulations outperform the accuracy of state-of-the-art macroscale approaches in terms of fiber orientation and fiber volume fraction.
However, it remains to be shown that they are able to predict the necessary compression forces considering non-isothermal, non-Newtonian and compaction behavior.
In this contribution, both approaches are applied to the elongational flow in a press rheometer and compared to experiments with 23\% glass fiber volume fraction.
The results show that both models predict contributions to the total compression force and orientation reasonably well for short flow paths. 
For long flow paths and thick stacks, complex deformation mechanisms arise and potential origins for deviation between simulations models and experimental observations are discussed. 
Furthermore, Jeffery's basic model is able to predict orientations similar to the high-fidelity mesoscale model.
For planar SMC flow, this basic model appears to be even better suited than the more advanced orientation models with diffusion terms developed for injection molding.
\end{abstract}



\begin{keyword}
Compression Molding \sep Sheet Molding Compound \sep Discontinuous Fiber Reinforcement
\end{keyword}

\end{frontmatter}

\section{Introduction}
Sheet molding compounds (SMC) are discontinuously reinforced polymer composites that are produced in a compression molding process.
The process allows cost-efficient production of complex parts because the material can flow in complex shapes and forms features, such as ribs, beads, or overmolded inserts.
The fibers are usually about \SI{25}{\milli\meter} long, which is much longer than in injection molding processes and leads to better mechanical performance. 
However, these long fibers pose challenges to simulation and modeling, even after decades of research on SMC compression molding dating back to the 1980s~\cite{SilvaNieto.1980, Tucker1983AFilling, Barone.1986}.

The first step in SMC manufacturing is the production of preimpregnated material on a SMC line. 
Two polymer foils are coated with a thermosetting matrix, fiber bundles are introduced between the foils and the resulting sandwich is coiled for storage.
A maturing processes increases the resin viscosity during storage, which enables further processing of the sheets by cutting and stacking.
An initial stack of SMC sheets at room temperature is then formed by compression molding in a heated mold.
This process generally involves heat transfer, curing of the thermosetting resin, suspension flow with reorientation of fiber bundles, fiber-matrix separation effects, weld-line formation, friction at the mold and the release of air trapped in voids of the initial stack. 
It is desirable to simulate such effects to account for them during mold design and use the results in subsequent structural simulations~\cite{Gorthofer2019, Romanenko2022AdvancedApplications}.

Early models describe the SMC flow with two-dimensional approaches, as SMC parts often have a planar shape.
\citeauthor{SilvaNieto.1980}~\cite{SilvaNieto.1980} assumed an isothermal Newtonian material without fiber reorientation and solved the flow based on a Poisson type equation for the pressure.
\citeauthor{Tucker1983AFilling}~\cite{Tucker1983AFilling} proposed a thickness averaged Hele-Shaw model that is solved by the finite element method (FEM) with the incorporation of heat transfer, non-Newtonian viscosity and curing. 
The authors advanced the Langrangian mesh with the flow front and remeshed the domain during the simulation.
They compared non-Newtonian isothermal simulations to experimental results and concluded that isothermal Newtonian models are limited to sufficiently thin parts~\cite{Lee.1984}.    
\citeauthor{Osswald1988}~\cite{Osswald1988} solved the two-dimensional compression molding problem on complex domains with finite elements and the boundary element method to mitigate the need of a finite element mesh~\cite{Osswald.1990}.
\citeauthor{Barone.1986}~\cite{Barone.1986} performed experiments with colored SMC sheets to analyze flow kinematics and observed that the flow rather resembles a \emph{plug-flow} instead of the parabolic profile with no-slip conditions at mold walls used in the previous models.
They propose a hydrodynamic friction model for the contact between SMC and mold, which is motivated by a small heated lubrication layer close to the mold surfaces.
Efforts to parameterize such plug-flow models and obtain correct compression forces were presented by several authors~\cite{Abrams.2003, Dumont.2003b, Dumont.2007, Hohberg.2017b, FerreSentis2021}.
A recent review on the numerical modeling of SMC compression molding is given by \citeauthor{Alnersson2020}~\cite{Alnersson2020}.

Fibers reorient during compression molding and models for the reorientation of short, rigid fibers are often based on Jeffery's equation~\cite{Jeffery.1922}. 
Fiber orientation tensors were introduced to simplify the description of multiple suspended fibers~\cite{Advani.1987} and several empirical parameters for fiber interaction~\cite{Folgar.1984}, reduced strain~\cite{Wang.2008} and anisotropic interaction~\cite{Phelps.2009} were introduced to enhance Jeffery's equation.
These models are successfully applied to injection molding applications, but generally the perquisites do not strictly apply for SMC. 
Thus, microscale models may be used to perform computational rheology experiments~\cite{LeCorre.2005}, to fit macroscopic model parameters~\cite{Dumont.2009, Guiraud.2012, Meyer2020a}, or to investigate critical molding areas~\cite{LondonoHurtado.2007, Kuhn2017}.
The simulation of all fibers in a component is computationally still unfeasible, but several authors~\cite{Guiraud.2012, Dumont.2007b, Le.2008, Motaghi.2019} observed that fibers typically stay in a bundled configuration during SMC compression molding.
A compression molding simulation on component scale with individual bundles at mesoscale was demonstrated in~\cite{Meyer2020} and showed accurate results of the fiber architecture, when compared to CT scans.
The mesoscale simulation accounts for anisotropic flow, varying fiber volume content and can predict fiber matrix segregation in confined regions~\cite{Rothenhausler2022ExperimentalApproaches}.
However, the previous work~\cite{Meyer2020, Meyer2021, Rothenhausler2022ExperimentalApproaches} simplified the rheology and focused on fiber architecture, but did not validate results of the computed pressures yet.

Therefore, this contribution aims at simulating a non-isothermal, non-Newtonian and compressible compression molding process on the mesoscale, i.e. resolving fiber bundles.
For reference, a one-dimensional macroscale model is formulated and solved.
The simulation results are compared to a flow in a rheological tool equipped with several pressure sensors along the flow path of SMC.

\section{Experiments}
The SMC under investigation is based on an unsaturated polyester-polyurethane hybrid (UPPH) resin that was developed to improve co-molding with unidirectional carbon fiber patches~\cite{Bucheler.2016}.
It has a glass fiber volume fraction of 23\% with \SI{25}{\milli\meter} fiber length.
The fibers are grouped in bundles with 200 fibers each in a Multistar 272 multi-end roving by Johns Manville.

\subsection{Transverse thermal properties}
\begin{figure}[!htpb]
    \centering
    \footnotesize
    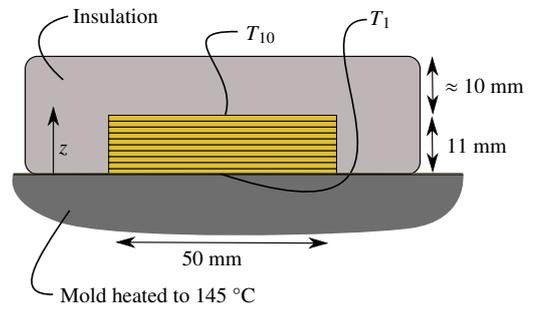
    \caption{Setup for the evaluation of the transverse heat conductivity: Ten sheets of SMC are stacked with temperature sensors (T\textsubscript{1} to T\textsubscript{10}) located centrally between each layer. 
    The stack is embedded in glass wool insulation and then placed on a heated steel plate.}
    \label{fig:temp_setup}
\end{figure}

Thermal properties of uncured UPPH GF-SMC are determined from temperature measurements in a stack of SMC sheets. 
Ten sheets of \SI{50}{\milli\meter} x \SI{50}{\milli\meter} x \SI{1.1}{\milli\meter} are stacked to a total stack height $H=\SI{11}{\milli\meter}$.
Thermocouples are positioned centrally between each layer pair during stacking and the so prepared stack is embedded in a glass wool insulation fitting the stack with sensors (see Figure \ref{fig:temp_setup}).
The first sensor T1 is located on the bottom surface of the stack, so that it is located between mold and stack when the stack is placed on a heated mold surface at \SI{145}{\celsius}.
A small weight on top of this configuration ensures proper contact.

\begin{figure}
        \centering
        \footnotesize
        \input{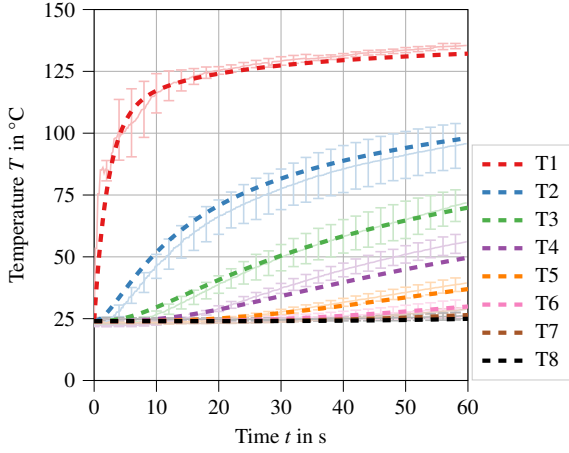}
        \caption{Measured temperatures (light solid lines with error bars indicating standard deviation) and best fit based on one-dimensional heat transfer equation (dark dashed lines) for sensors T1-T8. 
        Sensors T9 and T10 are excluded, as they do not detect any noticeable temperature change.}
        \label{fig:temp_result}
\end{figure}

The specific heat capacity $c_\textrm{p}$ is approximately \SI{1530}{\joule\per\kilo\gram\per\kelvin}, which is estimated from data of the resin~\cite{Schwab2019ReactionDynamics} and glass by rule of mixture.
The transient heat transfer is approximated as a one-dimensional process transverse to the sheets
\begin{equation}
    \rho c_\textrm{p} \frac{\partial T}{\partial t} = \kappa \frac{\partial^2 T}{\partial z^2} \quad z \in [0, H]
\end{equation}
with the boundary conditions
\begin{align}
    \kappa \frac{\partial T}{\partial z} &= 0                       & z=H\\
    \kappa \frac{\partial T}{\partial z} &= -k (T_\textrm{M}-T)     & z=0
\end{align}
as well as a constant mold temperature $T_\textrm{M}=\SI{145}{\celsius}$, the gap conductance $k$ and the thermal conductivity $\kappa$. 
Initially, the temperature $T$ is homogeneous at $T_0=\SI{24}{\celsius}$.
An optimal fit to the measured data is obtained by solving the initial boundary value problem for an initial guess of $\kappa$ and $k$, computing a scalar squared error and minimizing the error iteratively. 
The experimental results and the optimal fit are shown in Figure~\ref{fig:temp_result}.
The corresponding parameters are summarized in Table~\ref{tab:temp}.

\begin{table}
    \centering
    \footnotesize
    \begin{tabular}{lr}
        \toprule
        \textbf{Property} &  \textbf{Value} \\
        \midrule
        Thermal conductivity  $\kappa$ & \SI{0.163}{\watt\per\meter\per\celsius} \\
        Gap conductance  $k$ & \SI{403}{\watt\per\meter\squared\per\celsius} \\
        \bottomrule
    \end{tabular}
    \caption{Transverse thermal properties of UPPH-GF SMC in B-staged state.}
    \label{tab:temp}
\end{table}

\subsection{Viscosity of the SMC paste}
Specimens for viscosity measurement of the SMC paste are prepared by filling the paste in a mold instead of processing it on the SMC line. 
The mold is filled up to approximately \SI{1}{\milli\meter} thickness and sealed with styrene-tight foil.
The paste is matured for two weeks at room temperature similar to the SMC for molding.
Round coupons of \SI{25}{\milli\meter} diameter are cut from the thickened paste and placed in a Anton Paar MCR501 rheometer in plate-plate configuration.

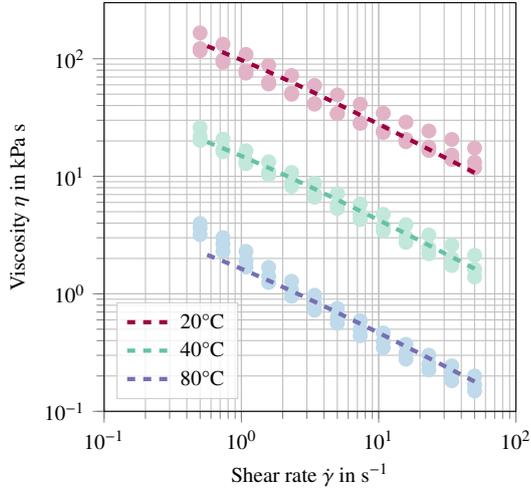
\begin{figure}[!htb]
    \centering
    \footnotesize
    \begin{tikzpicture}
\begin{axis}[
width=7cm,
height=7cm,
legend cell align={left},
legend style={
    at={(0.03,0.03)}, 
    anchor=south west,    
    draw=white!80!black,
    fill=white, 
    fill opacity=0.8, 
    draw opacity=1,
    text opacity=1},
log basis x={10},
log basis y={10},
tick align=outside,
tick pos=left,
xlabel={Shear rate $\dot{\gamma}$ in \si{\per\second}},
xmajorgrids,
xmin=0.1, xmax=100,
xminorgrids,
xmode=log,
ylabel={Viscosity $\eta$ in kPa s},
ymajorgrids,
ymin=0.1, ymax=300,
yminorgrids,
ymode=log,
clip mode=individual,
every axis plot/.append style={ultra thick}
]
\addplot+[Spectral-A!30!white, only marks, forget plot]
table {%
50 13.2
34.1 15.2
23.2 17.7
15.8 20.6
10.8 24.2
7.34 28.6
5 34.1
3.41 41.1
2.32 49.9
1.58 61
1.08 75.5
0.734 93.4
0.5 117
};
\addplot+[Spectral-A!30!white, only marks, forget plot]
table {%
50 11.8
34.1 13.9
23.2 16.6
15.8 19.7
10.8 23.6
7.34 28.2
5 34.2
3.41 41.6
2.32 51.2
1.58 63.4
1.08 79.1
0.734 97.6
0.5 122
};
\addplot+[Spectral-A!30!white, only marks, forget plot]
table {%
50 17.4
34.1 20.6
23.2 24.3
15.8 28.9
10.8 34.4
7.34 41.1
5 49.3
3.41 59.3
2.32 72.1
1.58 88.1
1.08 109
0.734 133
0.5 166
};
\addplot+[Spectral-A, dashed]
table {%
50 10.7191262190232
34.1 13.4856411359279
23.2 16.9657907285467
15.8 21.2921162119866
10.8 26.6037222420668
7.34 33.2580039034316
5 41.3770922621317
3.41 51.2271903844597
2.32 63.1813708808923
1.58 77.4030707980966
1.08 93.9629643917601
0.734 113.42596554654
0.5 135.433278284805
};
\addlegendentry{20°C}
\addplot [Spectral-L!40!white, only marks, forget plot]
table {%
50 1.62
34.1 2
23.2 2.47
15.8 3.05
10.8 3.77
7.34 4.66
5 5.75
3.41 7.1
2.32 8.8
1.58 10.9
1.08 13.6
0.734 16.9
0.5 21.7
};
\addplot [Spectral-L!40!white, only marks, forget plot]
table {%
50 1.39
34.1 1.74
23.2 2.19
15.8 2.74
10.8 3.43
7.34 4.28
5 5.32
3.41 6.63
2.32 8.23
1.58 10.3
1.08 12.8
0.734 16.2
0.5 20.2
};
\addplot [Spectral-L!40!white, only marks, forget plot]
table {%
50 2.13
34.1 2.6
23.2 3.17
15.8 3.88
10.8 4.75
7.34 5.82
5 7.13
3.41 8.76
2.32 10.8
1.58 13.3
1.08 16.5
0.734 20.7
0.5 26
};
\addplot+[Spectral-L, dashed]
table {%
50 1.63630157912951
34.1 2.05861703980415
23.2 2.58987062873035
15.8 3.25029509577483
10.8 4.06112511653804
7.34 5.07691794964602
5 6.31631720952745
3.41 7.81995946383189
2.32 9.64479518493737
1.58 11.8157734491105
1.08 14.3436828592487
0.734 17.3147589407716
0.5 20.6742305852152
};
\addlegendentry{40°C}
\addplot [Spectral-M!30!white, only marks, forget plot]
table {%
50 0.169
34.1 0.206
23.2 0.253
15.8 0.314
10.8 0.393
7.34 0.496
5 0.631
3.41 0.819
2.32 1.08
1.58 1.41
1.08 1.92
0.734 2.66
0.5 3.62
};
\addplot [Spectral-M!30!white, only marks, forget plot]
table {%
50 0.149
34.1 0.182
23.2 0.225
15.8 0.278
10.8 0.348
7.34 0.44
5 0.561
3.41 0.726
2.32 0.954
1.58 1.25
1.08 1.69
0.734 2.3
0.5 3.19
};
\addplot [Spectral-M!30!white, only marks, forget plot]
table {%
50 0.2
34.1 0.244
23.2 0.301
15.8 0.372
10.8 0.465
7.34 0.59
5 0.752
3.41 0.973
2.32 1.28
1.58 1.68
1.08 2.3
0.734 3.01
0.5 3.97
};
\addplot+[dashed]
table {%
50 0.179483036797758
34.1 0.225806075493873
23.2 0.284078345512037
15.8 0.356519141531889
10.8 0.445457657701257
7.34 0.556878183580483
5 0.692825704383313
3.41 0.857757573607298
2.32 1.05792058821203
1.58 1.29605136840904
1.08 1.57333329703897
0.734 1.89922539692474
0.5 2.26771992169411
};
\addlegendentry{80°C}
\end{axis}

\end{tikzpicture}
    \caption{Measured viscosities at different temperatures (light dots) and best fit to viscosity model (dark dashed lines) given in Equation~\eqref{eq:cross_wlf}.}
    \label{fig:viscosity_result}
\end{figure}

The viscosity as measured in oscillatory mode at \SI{20}{\celsius}, \SI{40}{\celsius} and \SI{80}{\celsius} for three specimens each. While the peroxide initiator starts to create free radicals at \SI{60}{\celsius}, significant fast cross-linking occurs typically at \SI{100}{\celsius} and the temperatures are chosen to cover the relevant temperature range for the uncured resin paste.
The measured viscosity shows typical power-law behavior.
As the power-law behavior has a singularity at zero shear rate, a Cross-WLF-like model
\begin{equation}
    \eta = \frac{\eta_0}{1+\left(\frac{\dot{\gamma}}{\dot{\gamma}_0}\right)^{1-n}}
    \quad 
    \textrm{with}
    \quad 
    \eta_0 = D_1 e^{\frac{-\alpha_1(T - T^*)}{\alpha_2 + (T - T^*)}}
    \label{eq:cross_wlf}
\end{equation}
and with a fixed transition shear rate $\dot{\gamma}_0=0.1$ is employed to limit the viscosity at small shear rates. The power-law coefficient $n$, transition shear rate $\gamma_0$ and parameters $T^*$, $D_1$, $\alpha_1$, $\alpha_2$ are fitted to the experimental results. 
These parameters should not be considered strictly as Cross-WLF parameters, as there is no plateau and the model is mainly chosen to smoothly transition to a finite viscosity as the shear rate approaches zero.
The fit is obtained by minimizing the sum of squares of the normalized least-squares error at each measured temperature for all specimens.
The experimental results and the optimal fit is shown in Figure~\ref{fig:viscosity_result}. 
The corresponding parameters are summarized in  Table~\ref{tab:viscosity}.
    
\begin{table}
    \centering
    \footnotesize
    \begin{tabular}{lr}
        \toprule
        \textbf{Property} & \textbf{Value} \\
        \midrule
        Reference viscosity $D_1$               & \SI{72}{\kilo\pascal\second}\\
        Transition shear rate $\dot{\gamma_0}$  & 0.1\\
        Power law coefficient $n$               & 0.385\\
        Glass transition temperature $T^*$      & \SI{40.73}{\celsius}\\
        Fitting parameter $\alpha_1$            & 7.94\\
        Fitting parameter $\alpha_2$            & \SI{105.96}{\celsius}\\      
        \bottomrule
    \end{tabular}
    \caption{Viscous properties of UPPH paste in B-staged state.}
    \label{tab:viscosity}
\end{table}

\subsection{Press rheometer trials}
\begin{figure}[!ht]
    \centering
    \footnotesize
    \begin{tikzpicture}

\begin{axis}[
hide x axis,
hide y axis,
tick align=outside,
tick pos=left,
width=10cm,
xmin=-40, xmax=840,
xtick distance={2.0},
ymin=-300, ymax=300,
axis equal image,
]
\path [draw=white!50!black, fill=white!50!black, opacity=0.2,rounded corners=2]
(0,-225) rectangle (200,225);


\path [draw=white!50!black, fill=white!50!black, opacity=0.2, rounded corners=2]
(0,-225) rectangle (600,225);

\draw [draw=white!50!black, fill=white!50!black, opacity=0.2, rounded corners=2]
(0,-225) rectangle (800,225);
\node[] at (axis cs: 150,-100) {25\%};
\node[] at (axis cs: 550,-100) {75\%};
\node[] at (axis cs: 750,-100) {100\%};

\draw [draw=black, rounded corners=2] (0,-225) rectangle (800,225);

\draw[thick,-stealth] (0,0) -- (100,0) node[midway, below] {$x$};
\draw[thick,-stealth] (0,0) -- (0,100) node[midway, left] {$y$};
\addplot [black, mark=*, mark size=0.5]
table {%
0 0
};

\node[anchor=south] at (32,0) {P\textsubscript{1}};
\addplot [Dark2-A, mark=*, mark size=2.0, line width=1.0, mark options={solid,fill opacity=0.5}]
table {%
32 0
};
\node[anchor=south] at (146,0) {P\textsubscript{2}};
\addplot [Dark2-B, mark=*, mark size=2.0, line width=1.0, mark options={solid,fill opacity=0.5}]
table {%
146 0
};
\node[anchor=south] at (248,0) {P\textsubscript{3}};
\addplot [Dark2-C, mark=*, mark size=2.0, line width=1.0, mark options={solid,fill opacity=0.5}]
table {%
248 0
};
\addplot [white!50!black, mark=o, mark size=2.0, line width=1.0, mark options={solid,fill opacity=0}]
table {%
350 0
};
\node[anchor=south] at (450,0) {P\textsubscript{5}};
\addplot [Dark2-E, mark=*, mark size=2.0, line width=1.0, mark options={solid,fill opacity=0.5}]
table {%
450 0
};
\node[anchor=south] at (552,0) {P\textsubscript{6}};
\addplot [Dark2-F, mark=*, mark size=2.0, line width=1.0, mark options={solid,fill opacity=0.5}]
table {%
552 0
};
\node[anchor=south] at (604,0) {P\textsubscript{7}};
\addplot [Dark2-G, mark=*, mark size=2.0, line width=1.0, mark options={solid,fill opacity=0.5}]
table {%
604 0
};
\addplot [white!50!black, mark=o, mark size=2.0, line width=1.0, mark options={solid,fill opacity=0}]
table {%
654 0
};
\node[anchor=south] at (709,0) {P\textsubscript{9}};
\addplot [Dark2-H, mark=*, mark size=2.0, line width=1.0, mark options={solid,fill opacity=0.5}]
table {%
709 0
};
\addplot [white!50!black, mark=o, mark size=2.0, line width=1.0, mark options={solid,fill opacity=0}]
table {%
764 0
};
\addplot [white!50!black, mark=o, mark size=2.0, line width=1.0, mark options={solid,fill opacity=0}]
table {%
500 165
};
\addplot [white!50!black, mark=o, mark size=2.0, line width=1.0, mark options={solid,fill opacity=0}]
table {%
500 -165
};
\addplot [white!50!black, mark=o, mark size=2.0, line width=1.0, mark options={solid,fill opacity=0}]
table {%
500 0
};
\end{axis}

\end{tikzpicture}
    \caption{Press rheometer (\SI{800}{\milli\meter} x \SI{450}{\milli\meter}) with pressure sensor locations. Active pressure sensor locations are color-coded and labeled P\textsubscript{s}, alternate mounting positions are drawn as gray circles.}
    \label{fig:press_rheometer}
\end{figure}
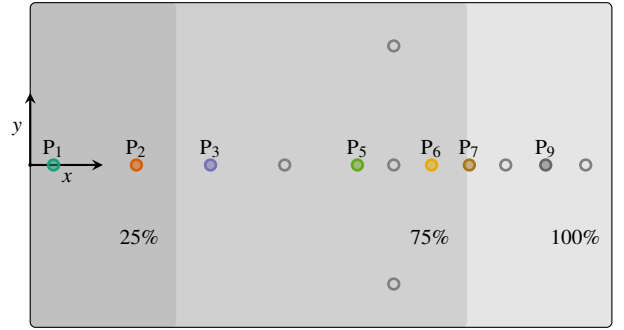

Compression trials are performed using a tool with a length $L$ of \SI{800}{\milli\meter} and a width $W$ of \SI{450}{\milli\meter}. 
The initial stacks always fill the entire width and are aligned to one end of the mold, while the length (and therefore mold coverage) varies between 25\% and 100\% (see gray regions in Figure~\ref{fig:press_rheometer}).
The mold is equipped with several pressure sensors (6167ASP by Kistler Instrumente GmbH, Sindelfingen, Germany) along the flow direction.
The color-coded locations in Figure~\ref{fig:press_rheometer} are active during the trials in this work.
The mold is heated to \SI{145}{\celsius} and mounted to a hydraulic press with parallelism control (COMPRESS PLUS DCP-G 3600/3200 AS by Dieffenbacher GmbH Maschinen- und Anlagenbau, Eppingen, Germany). 
All trials are performed with a constant closing speed of $\dot{h}$ = -\SI{1}{\milli\meter\per\second} and a maximum compression force $F_\textrm{max}=\SI{4400}{\kilo\newton}$. 

\subsubsection{Hydrodynamic mold friction}
\begin{figure}[!ht]
    \centering
    \footnotesize
    \input{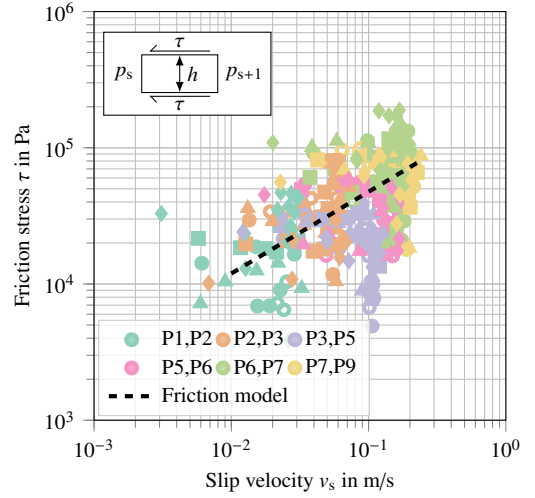}
    \caption{
    Friction stress evaluation. 
    Colors refer to the first sensor of each sensor pair according to Figure~\ref{fig:press_rheometer}. 
    The marker styles differentiate individual molding trials with 75\% initial mold coverage each. 
    The hydrodynamic power-law friction model with parameters from Table~\ref{tab:friction} is indicated by the dashed black line.
    }
    \label{fig:friction}
\end{figure}

The press rheometer enables a characterization of the hydrodynamic friction between SMC and the molds caused by the lubrication layer. 
Assuming an ideal incompressible plug-flow in the tool, i.e. $\partial v_x / \partial z = 0$, friction stresses can be computed from the pressure differences between sensor pairs. 
The force balance for SMC between two sensors at position $\Delta x_\textrm{s}$ with pressure $p_\textrm{s}$ and position $\Delta x_\textrm{s+1}$ with pressure $p_\textrm{s+1}$ reads
\begin{equation}
    2 \tau \left(x_\textrm{s+1} -x_\textrm{s}\right) W = h W \left(p_\textrm{s+1} - p_\textrm{s}\right)
\end{equation}
with the average friction shear stress $\tau$ and the current thickness of SMC $h$ (see inset of Figure~\ref{fig:friction}).
Therefore 
\begin{equation}
    \tau = \frac{h \left(p_\textrm{s+1} - p_\textrm{s}\right)}{2 \left(x_\textrm{s+1} -x_\textrm{s}\right)}
\end{equation}
holds.
The corresponding slip velocity $v_\textrm{s}$ at a mid point between two sensors can be approximated from the continuity equation as  
\begin{equation}
    v_\textrm{s} = \frac{\dot{h}}{h} \left(x_\textrm{s} + \frac{\left(x_\textrm{s+1} -x_\textrm{s}\right)}{2}\right) 
\end{equation}
A commonly used relation between shear stress and the slip velocity is a hydrodynamic power-law model 
\begin{equation}
    \label{eq:friction}
    \tau = -\lambda \left( \frac{v_\textrm{s} }{v_0}\right)^{m-1} v_\textrm{s}
\end{equation}
with a power-law coefficient $m$, a hydrodynamic friction coefficient $\lambda$ and an arbitrary reference velocity $v_0$ to normalize the power-law term~\cite{Dumont.2007, Hohberg.2017b}.
The characterization of these parameters is notoriously difficult and subjected to strong uncertainties, as they can be only obtained by large scale in-mold rheology.
Due to high pressures, uncertainties of pressure sensors and deviations from the assumed ideal flow kinematics, evaluation of friction stress parameters is ambiguous. 
To mitigate some of uncertainty, the evaluation considers only pressure differences $p_\textrm{s} - p_\textrm{s+1} > \SI{5}{\bar}$ due to the low precision of sensors.
Nonetheless, the choice of parameters given in Table~\ref{tab:friction}, which are typical values according to relevant literature, give a reasonable parameterization of shear stress featuring a power law behavior (see Figure~\ref{fig:friction}).
The linear trend towards higher friction stresses with higher slip velocities in the double logarithmic diagram n Figure~\ref{fig:friction} supports the application of a power-law friction model.

\begin{table}[!ht]
    \centering
    \footnotesize
    \begin{tabular}{lr}
        \toprule
        \textbf{Property}                               & \textbf{Value} \\
        \midrule
        Reference velocity $v_0$                        &  \SI{1}{\milli\meter\per\second} \\
        Power-law coefficient $m$                       & 0.6 \\
        Hydrodynamic friction coefficient $\lambda$     & \SI{3.0}{\mega\newton\second\per\meter\cubed} \\
        \bottomrule
    \end{tabular}
    \caption{Mold friction parameters}
    \label{tab:friction}
\end{table}

\subsubsection{Compaction behavior}
The SMC under investigation shows compressible behavior, as previously reported in~\cite{Hohberg.2017b}.
The compressibility in recent high-performance SMCs is caused by their high initial pore content~\cite{FerreSentis2017}. 
Specimens with 100\% mold coverage are consolidated under typical SMC process conditions to determine an equation of state that relates pressure and volumetric strain.
Automated cutting of SMC sheets ensures an accurate initial mold coverage and compression data (force, mold displacements) is recorded during the trials.

\begin{figure}[!ht]
    \centering
    \footnotesize
    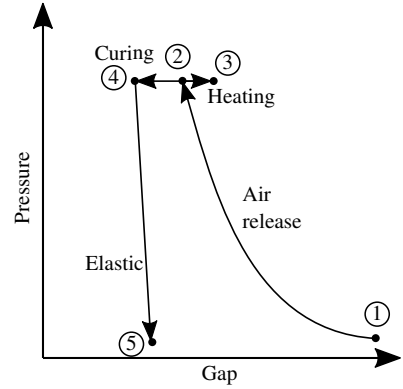
    \caption[Schematic compressibility]{
      Schematic relation between pressure and mold gap for compression.
      1) First contact between mold and SMC stack.
      2) The maximum compression force is reached.
      3) The maximum thermal extension is reached.
      4) The part is fully cured.
      5) The part is demolded.
    }
    \label{fig:compression_sketch}
\end{figure}
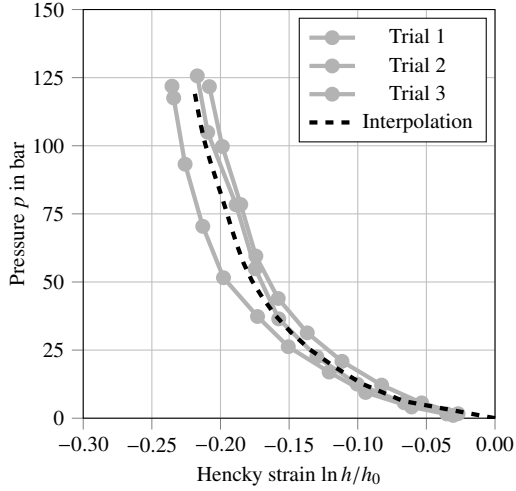
\begin{figure}[!ht]
    \centering
    \footnotesize
    \begin{tikzpicture}
\begin{axis}[
width=7cm,
height=7cm,
clip mode=individual,
tick align=outside,
tick pos=left,
xlabel={Hencky strain $\ln{h/h_0}$},
xmajorgrids,
xmin=-0.3, 
xmax=0,
xtick={-0.3, -0.25, -0.2, -0.15, -0.1, -0.05, 0.0},
xticklabel style={
        /pgf/number format/fixed,
        /pgf/number format/fixed zerofill,
        /pgf/number format/precision=2
},
ylabel={Pressure $p$ in \si{\bar}},
ylabel shift=-1.5mm,
ymajorgrids,
ymin=0, 
ymax=150,
ytick={0, 25, 50, 75, 100, 125, 150},
every axis plot/.append style={ultra thick}
]
\addplot [black!30!white, mark=*]
table {%
-0.0304589084654714  1.02
-0.0608097590195937  4.12
-0.0942791640452611  9.45
-0.120826497517542  16.93
-0.150615078411869  26.24
-0.173021068406502  37.33
-0.197814000076862  51.56
-0.212891460959272  70.41
-0.225802230878571  93.23
-0.234105631598876 117.55
-0.235330528861762 121.87
};
\addlegendentry{Trial 1}
\addplot [black!30!white, mark=*]
table {%
-0.0350984998057698 1.55
-0.0661422479238409 5.66
-0.100393488418309 12.51
-0.129815312776008 22.52
-0.157561237789788 36.46
-0.174669079569086 54.84
-0.188643719984357 78.31
-0.20933900476223  104.94
-0.216915338616155 125.65
};
\addlegendentry{Trial 2}
\addplot [black!30!white, mark=*]
table {%
-0.0269313432645571 1.65
-0.0534216677538994 5.63
-0.0826086068531238 12.15
-0.111490359922559 20.88
-0.136788079693066 31.33
-0.157962926826262 43.93
-0.174247785757456 59.54
-0.185243068434111 78.42
-0.198641373542458 99.74
-0.208079617791534 121.71
};
\addlegendentry{Trial 3}
\addplot+[black, dashed]
table {%
0 0
-0.0668219261650857  6.31
-0.0968282238917516 12.63
-0.117220076601342  18.94
-0.134943900566671  25.26
-0.148811088778482  31.57
-0.160234457860114  37.89
-0.169341204711059  44.21
-0.177165332780559  50.52
-0.183107377661     56.84
-0.187686679905568  63.15
-0.191850448117144  69.47
-0.195594215327905  75.78
-0.199551267178133  82.10
-0.203701935563667  88.42
-0.207739960006085  94.73
-0.211331222715002 101.05
-0.214257567026267 107.36
-0.216650419391886 113.68
-0.218995725976169 120.00
};
\addlegendentry{Interpolation}
\end{axis}
\end{tikzpicture}
    \caption{Compaction behavior of the UPPH-GF SMC.
    The recorded data is corrected with the stiffness of the mold and press, which is determined from an empty stroke. The corrected data is shown in light colors.
    The rising flank is interpolated and used as tabulated data to describe the relation between Hencky strain and pressure.}
    \label{fig:compression}
\end{figure}

The expected behavior is illustrated in the schematic in Figure~\ref{fig:compression_sketch}: 
Upon first contact, the bulk material offers small resistance to compression, as trapped air is compressed and released. 
The resistance increases, as an increasing amount of air pockets is closed and air escapes until the maximum compression force is reached. 
At this constant compression force, the material first expands due to heating and consequently shrinks due to cross linking of the thermosetting polymer. 
Finally, the cured part expands elastically as the compression force is relaxed and the mold opens.
The final recorded gap is the part thickness.

Only the region of the rising flank between point 1 and 2 is of interest for the compression molding simulation here. 
Therefore, these points are extracted and plotted over the Hencky strain $E=\ln(h/h_0)$ in Figure~\ref{fig:compression}.
The resulting high compressibility is comparable to other structural SMCs reported in recent literature~\cite{FerreSentis2021}.
An averaged relation between strain and pressure is obtained by averaging the strains at various pressure levels.
This tabulated data is then used to interpolate the equation of state $p(E)$ in subsequent simulations.

\section{Models}
\subsection{Governing equations}
The governing equations for the SMC deformation in a domain $\Omega$ are the conservation of mass, conservation of momentum and conservation of internal energy
\begin{align}
	\label{eq:mass_balance_3d}
	\frac{\partial \rho}{\partial t} + \nabla \cdot (\rho \bm{v}) &= 0 &\bm{x} \in \Omega\\
	\label{eq:momentum_balance_3d}
	\frac{\partial (\rho \bm{v})}{\partial t} + \nabla \cdot \left((\rho \bm{v})  \otimes \bm{v} \right) &= \nabla \cdot \bm{\sigma}  + \bm{f} &\bm{x} \in \Omega\\
	\label{eq:energy_balance_3d}
	\frac{\partial (\rho c_\textrm{p}T)}{\partial t} + \nabla \cdot \left( (\rho c_\textrm{p}T) \bm{v}\right) &= \nabla \cdot \bm{h} &\bm{x} \in \Omega
\end{align}
with mass density $\rho \in \mathcal{R}$,  fluid velocity $\bm{v} \in \mathcal{R}^3$, Cauchy stress $\bm{\sigma} \in \mathcal{R}^{3 \times 3}$,  symmetric strain rate $\bm{D} \in \mathcal{R}^{3 \times 3}$, a body force field $\bm{f} \in \mathcal{R}^3$, temperature $T \in \mathcal{R}$ and the non-convective heat flux $\bm{h} \in \mathcal{R}^3$.
The specific heat capacity $c_\textrm{p}$ is assumed to be constant and gravity as well as heating by viscous dissipation are neglected, as their influence on results is expected to be small.
The boundary of $\Omega$ is divided into the mold contact area $\partial \Omega_\textrm{M}$ and the free flow front $\partial \Omega_\textrm{F}$ with boundary conditions
\begin{align}
    \label{eq:bc_m}
    \bm{v} \cdot \bm{n} &= v_\textrm{M} & \bm{\sigma} \cdot \bm{n} &= \bm{\tau}    & \bm{h} \cdot \bm{n} &= - k \left(T_\textrm{M} - T\right)  & \bm{x} \in \partial \Omega_\textrm{M} \\
                        &               & \bm{\sigma} \cdot \bm{n} &= \bm{0}       & \bm{h} \cdot \bm{n} &= 0                                  & \bm{x} \in \partial \Omega_\textrm{F}, 
\end{align}
where $\bm{n}$ denotes the normal of a boundary surface and $v_\textrm{M}$ denotes the mold velocity ($v_\textrm{M}=0$ at bottom, $v_\textrm{M} = \dot{h}$ at top).
In this work, these equations are solved at mesoscale utilizing a three-dimensional model simulating the motion of individual fiber bundles and at macroscale using a reduced one-dimensional model for the elongational flow in a press rheometer. 

\subsection{Mesoscale direct bundle simulation}
\label{sec:dbs}
\subsubsection{Overview}
The basic idea of the Direct Bundle Simulation is the description of fiber bundles by chains of 1D truss elements that interact with matrix material during the compression molding process.
The trusses are subjected to hydrodynamic interaction forces by the matrix fluid, while the fluid experiences opposing forces. 
The direct simulation of bundles eliminates the need for fiber orientation models, closure approximations, modeling of long-range hydrodynamic interactions and improves the accuracy of simulated fiber architecture in regions, where scale-separation does not apply~\cite{Meyer2020}.

An operator splitting scheme is used to solve the governing Equations~\eqref{eq:mass_balance_3d},~\eqref{eq:momentum_balance_3d} and~\eqref{eq:energy_balance_3d} in a Coupled Eulerian-Lagrangian simulation framework implemented in SIMULIA Abaqus.
The method assigns an element volume fraction of material to each element and reconstructs the material surface that may interact with Lagrangian bodies representing molds at $\partial \Omega_\textrm{M}$~\cite{Benson.2004} or form a free surface at $\partial \Omega_\textrm{F}$.
The interaction between bundles and matrix is integrated via several user subroutines to the simulation, which define a body force field $\bm{f}$ based on the hydrodynamic interactions with suspended truss elements.

\subsubsection{Hydrodynamic interactions between bundles and matrix}
The motion of fiber bundle segments is governed by hydrodynamic forces acting between fluid and bundle segment as well as short range interactions with other bundles. The hydrodynamic force on a bundle segment $j$ is computed as 
\begin{equation}
\bm{F}_j = 6 \pi \eta R \left( k_\textrm{d} \Delta \bm{v}_j + k_\textrm{l} \left\lVert \Delta \bm{v}_j  \right\rVert \bm{q} \right) 
\label{eq:stokes_generalization}
\end{equation}
with matrix viscosity $\eta$, bundle radius $R$ and a direction $\bm{q}$ orthogonal to the bundle axis. 
The factors $k_\textrm{d}$ and $k_\textrm{l}$ are coefficients determined from micro-simulations and depend on segment aspect ratio as well as orientation of the bundle~\cite{Meyer2020}.
The matrix viscosity $\eta$ depends on the local temperature and shear rate of the neighboring fluid as defined in equation \eqref{eq:cross_wlf}, but is assumed locally constant at microscale (i.e. the coefficients $k_\textrm{d}$ and $k_\textrm{l}$ remain the same as for the Newtonian case).

The computation of the relative velocity $\Delta \bm{v}$ requires knowledge of the neighborhood relation between bundle segments and surrounding fluid cells.
The formal task is finding the set 
\begin{equation}
    \mathcal{S}_j = \{i \in \mathcal{E} \quad | \quad \| x_i - x_j \| < L\} 
\end{equation}
for each bundle center $x_j$, where $\mathcal{E}$ is the set of fluid cells defined by unique integer labels, $L$ is the search radius (typically equal to segment length) and $x_i, x_j \in \Omega$. 
A simplistic approach for this task is searching all neighbors for each bundle segment during each time step. 
This original formulation used in~\cite{Meyer2020} leads to a quadratic search complexity $\mathcal{O}(n^2)$ and slows down larger simulations quite significantly. 
To address this major bottleneck, the neighborhood search now utilizes a binary space partitioning tree (kd-tree). 
This improves the search complexity to $\mathcal{O}(n \log{n})$ and is implemented with the highly optimized library \textsc{KDTREE2}~\cite{Kennel2004X}.

The relative velocity is then determined from the neighbors as
\begin{equation}
	\Delta \bm{v}_j = \sum_{i \in S_j}  \frac{w_{ij}}{W_j} \left(\bm{v}_i-\bm{v}_j \right)
	\label{eq:relative_velocity}
\end{equation}
with Gaussian weighting factors $w_{ij}$ and $W_j=\sum_{i \in S} w_{ij}$.
The weighted hydrodynamic forces are used to apply an opposed body force field $\bm{f}$ on the fluid phase by summing up the contribution of each bundle $j$ in fluid element $i$
\begin{equation}
	\bm{f}_{ij} = -\frac{1}{V_i} \frac{w_{ij}}{W_j} \bm{F}_j
	\label{eq:bundle_body_force}
\end{equation} 
with the volume $V_i$ of the $i$-th element.

Fiber bundles in close proximity experience normal forces and lubrication forces due to the thin sheared fluid layer between them. Friction forces and lubrication moments are typically neglected for bundled suspensions~\cite{Servais1999, LeCorre.2005}.
While normal penetration of fiber bundles is prohibited by a linear contact stiffness, the tangential contact stresses are defined as 
\begin{equation}
\bm{\sigma}_\textrm{c} = 
-\eta \frac{d_a^2}{|\sin{\phi}|} \frac{\Delta \bm{v}_\textrm{t}}{\bar{t}(g) A}, g > 0
\label{eq:tangential_stress}
\end{equation}
with bundle width $d_a$, contact angle $\phi$, relative tangential velocity $\Delta \bm{v}_\textrm{t}$, contact area $A$ as well as the relation between physical gap $g$ and the effective sheared gap $\bar{t}(g)$~\cite{Meyer2021}. 
However, the tangential stresses are unbound and therefore can lead to infeasible small time steps. 
Therefore, the maximum tangential stress is bound to an upper limit such that it does not negatively affect the time step.

The presence of fiber bundles and the body force field $\bm{f}$ renders a generally anisotropic flow behavior of the SMC. 
It has been proven to agree with analytical solutions in simple compression for Newtonian matrix and absence of friction~\cite{Meyer2021}.

\subsubsection{Application to a press rheometer}

\begin{figure*}
    \centering
    \footnotesize
    \includegraphics[width=1.0\textwidth]{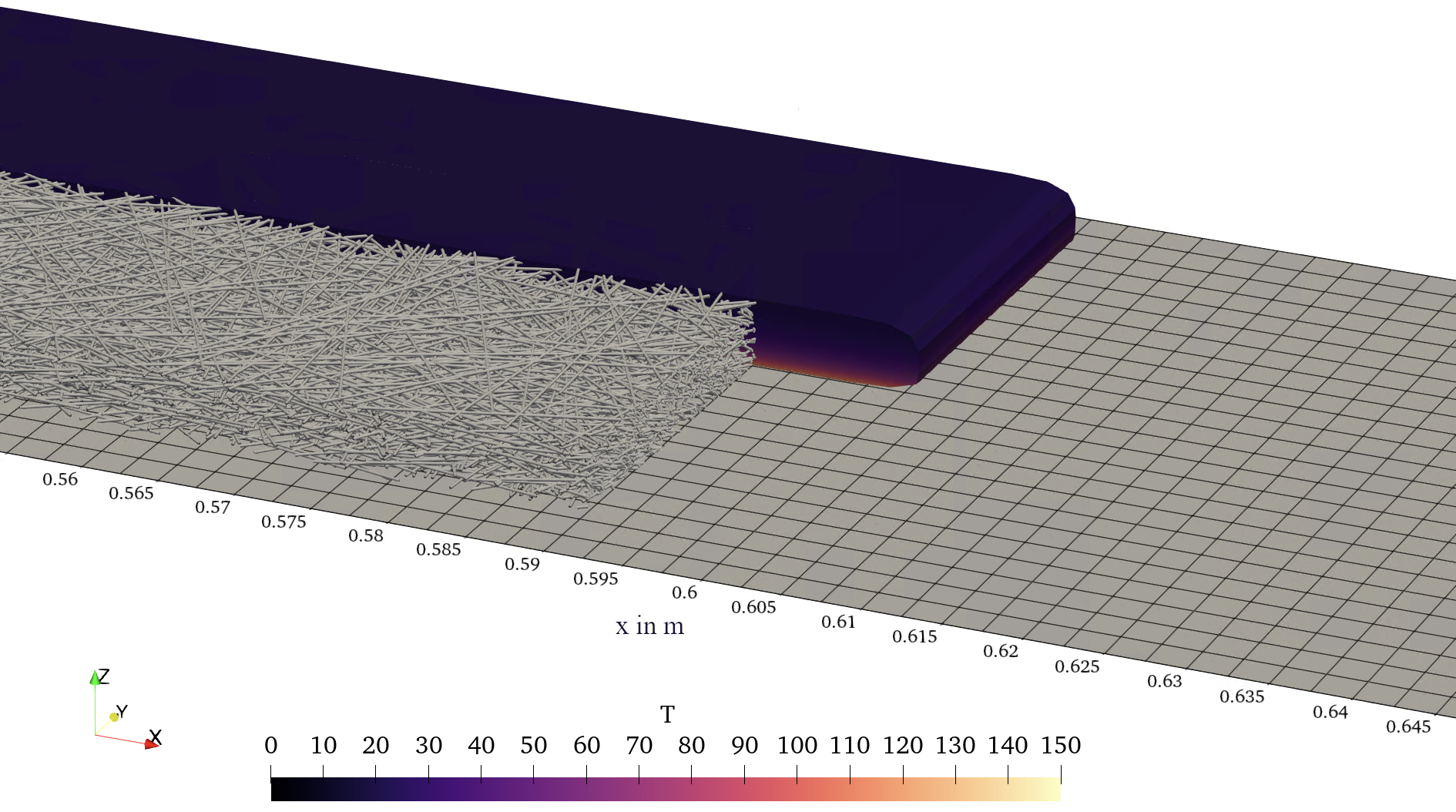}
    \caption{Initial setup of the mesoscale simulation model. The upper mold is not shown and cuts are applied to the filled Eulerian domain as well as the bundle structure for visualization purposes.}
    \label{fig:dbs_setup}
\end{figure*}

As the flow in a press rheometer is mostly elongational with no changes in the perpendicular direction, only a \SI{50}{\milli\meter} wide strip of the tool is modeled. 
The tool itself is represented by isothermal rigid bodies with a temperature of \SI{145}{\celsius}.
The rigid bodies are in contact with the SMC surface employing a penalty term to prevent penetration, a friction term implementing Equation~\eqref{eq:friction} and a thermal heat flux according to Equation~\eqref{eq:bc_m}.
The cavity is represented by an Eulerian domain, in which the initially filled subset $\Omega_0$ representing the SMC stack is defined with an element volume fraction field. 
In reality, the stack is in contact with the bottom mold for approximately \SI{10}{\second} until the upper mold arrives at the top of the stack. 
This yields to an unsymmetrical initial heat distribution, which is introduced to the simulation by solving the heat transfer process separately and applying the solution as initial temperature field.
Fiber bundles are generated within the stack $\Omega_0$ randomly from a uniform planar isotropic orientation distribution until the desired volume fraction of 23\% is reached. 
The bundles are \SI{25}{\milli\meter} long, are meshed with \SI{2.5}{\milli\meter} long truss elements and have a cross sectional area of \SI{0.03}{\milli\meter\squared}. 
The truss sections are elastic with $E=\SI{72}{\giga\pascal}$.
Bundle elements extending outside the stack domain are removed, such that shorter fibers are present close to the edges of a stack just like in the physical process.
The initial setup is shown in Figure~\ref{fig:dbs_setup}.

A global mass scaling factor of $10^4$ is applied to reduce computation time of the explicit time stepping procedure and it was verified that the kinetic energy due to this modification is small compared to the total work energy of compression.

\subsection{A one-dimensional macroscale reference model}
\subsubsection{Overview}
The mesoscale simulation model is applicable to arbitrary three-dimensional flows and has previously proven to provide accurate predictions of the fiber bundle architecture~\cite{Meyer2020, Rothenhausler2022ExperimentalApproaches}. 
However, its evaluation is computationally expensive and efficient macroscale models remain relevant for the application in simple planar SMC flows. 
Hence, a macroscale reference model for the elongational flow in a press rheometer is formulated as a one-dimensional model for comparison with the high fidelity mesoscale model. 

The macroscale reference model uses fiber orientation tensors to describe the orientation state as a stochastic moment of the fiber orientation distribution function $\psi : \mathcal{K}^2 \rightarrow \mathcal{R}_{\geq0}$, where $\psi$ describes the likelihood to find a fiber in a given direction $\bm{p} \in \mathcal{K}^2$. 
The second and fourth order fiber orientation tensors are defined here as 
\begin{align}
    \bm{A}      &= \int_{\mathcal{K}^2} \psi  (\bm{p}) \bm{p} \otimes \bm{p} \textrm{d} a \in \mathcal{R}^{3 \times 3}\\
    \mathbb{A}  &= \int_{\mathcal{K}^2} \psi  (\bm{p}) \bm{p} \otimes \bm{p} \otimes \bm{p} \otimes \bm{p} \textrm{d} a \in \mathcal{R}^{3 \times 3 \times 3 \times 3},
\end{align}
where $\int_{\mathcal{K}^2} \bullet \textrm{d}a$ denotes an integration over a unit sphere $\mathcal{K}^2$. 
Jeffery's equation~\cite{Jeffery.1922} can be expressed in terms of fiber orientation tensors as 
\begin{equation}
    \bm{\dot{A}} = \bm{W} \cdot \bm{A} + \bm{A} \cdot \bm{W} + \xi \left(\bm{D} \cdot \bm{A} + \bm{A} \cdot \bm{D} - 2 \mathbb{A} : \bm{D} \right)
\end{equation}
with a shape factor $\xi \in [0,1]$ and with $\dot{(\bullet)}$ denoting a material derivative.
The tensors $\bm{D} \in \mathcal{R}^{3 \times 3}$ and $\bm{W} \in \mathcal{R}^{3 \times 3}$ are the symmetric strain rate tensor and vorticity tensor, respectively.

The following simplifications are introduced for the one-dimensional reference model: 
\begin{itemize}
    \item There is no velocity gradient perpendicular to the flow ($D_{yy}=0$).
    \item The velocity gradient in compression direction is prescribed by the mold closing speed ($D_{zz}=\frac{\dot{h}}{h}$).
    \item The material undergoes a perfect plug-flow without shear ($D_{ij}=0 , i \neq j$ and $\bm{W} \equiv \bm{0}$).
    \item The fiber orientation state is planar ($A_{xz}=A_{yz}=A_{zz}=0$).
    \item The fibers are long and slender ($\xi=1$).
    \item An affine map $m: \Omega_\textrm{1D} \rightarrow \Omega_\textrm{1D}^*$ transforms the local coordinate $x \in [0, X_\textrm{max}]$ to a stretched coordinate $x^*=m(x)=x/X(t) \in [0,1]$. Here, $X \in [X_0, X_\textrm{max}]$ denotes the flow front position. Applying chain rule, the spatial gradient becomes $\frac{\partial \bullet}{\partial x} = \frac{\partial \bullet}{\partial x^*} \frac{\partial x^*}{\partial x}=\frac{1}{X}\frac{\partial \bullet}{\partial x^*}$.
    \item The solution variables are thickness averaged values depending only on time $t$ and one-dimensional position $x^*$.
\end{itemize}

\subsubsection{Average temperature}
The temperature between two closing plates with constant closing speeds and ideal heat transfer at the mold surfaces (Dirichlet boundaries) has been reported in literature~\cite{Lee.1981, Castro.1990}. 
Averaging this result for the temperature distribution over the thickness yields

\begin{equation}
    \bar{T}(t)
    = T_\textrm{M} + (T_\textrm{M}-T_0)\left[
        \frac{4}{\pi^2} \sum_{q=1}^\infty \frac{\cos(q \pi-1)}{q^2}
            \exp \left( \frac{-q^2\pi^2\kappa t}{h_0 h \rho c_\textrm{p}}\right)
    \right]
\end{equation}
as an approximation of the average temperature in the SMC material.
This explicit formulation depends only on time and can be evaluated separately.

\subsubsection{Constitutive model}
The SMC is modeled as compressible anisotropic viscous material.
Hence, the stress is computed as 
\begin{equation}
    \bm{\sigma} = -p(\rho) \bm{I} + \mathbb{V} : \bm{D}',
\end{equation}
where $p(\rho)$ is evaluated from the tabulated data shown in Figure~\ref{fig:compression}, $\bm{I}$ denotes the identity tensor and $\bm{D}'=\mathbb{P}_2 \bm{D}$ denotes the deviatoric strain rate tensor.\footnote{The identity tensor on symmetric fourth order tensors may be decomposed in a spherical and deviatoric projector $\mathbb{I}^\textrm{s}=\mathbb{P}_1+\mathbb{P}_2$. While the spherical projector $\mathbb{P}_1 = \frac{1}{3} \bm{I} \otimes \bm{I}$ can be used to extract the volumetric part of a second order tensor (such as strain rate tensor $\bm{D}$), the deviatoric projector $\mathbb{P}_2 = \mathbb{I}^\textrm{s}-\mathbb{P}_1$ may be used to obtain the deviatoric part.}
The anisotropic viscosity tensor is given by 
\begin{equation}
    \mathbb{V} = 2\eta \mathbb{P}_2 + \eta_\textrm{2} \left(\mathbb{A} - \frac{1}{3} \bm{I} \otimes \bm{A}\right) 
\end{equation}
with
\begin{equation}
    \eta_\textrm{2} = \frac{4 f r_\textrm{p}^2}{3 \left[ \ln (1/f) + \ln\ln (1/f) + C \right]} \eta
\end{equation}
for a semi-dilute suspension of rods with fiber volume fraction $f$, fiber aspect ratio $r_\textrm{p}$ and a constant parameter $C=0.1585$~\cite{Shaqfeh.1990}. 
The fourth-order orientation tensor $\mathbb{A}$ is computed with an IBOF closure approximation~\cite{DuChung.2002}.
Assuming planar orientation, the absence of strain perpendicular to the flow  and perfect plug-flow behavior, the expression reduces to only four non-trivial components 
\begin{align}
    V_\textrm{xxxx} &= \frac{4}{3} \eta + \eta_2 \left(A_\textrm{xxxx} - \frac{1}{3} A_\textrm{xx}\right)\\
    V_\textrm{zzxx} &= -\frac{2}{3} \eta - \frac{1}{3} \eta_2 A_\textrm{xx}\\
    V_\textrm{zzzz} &= \frac{4}{3} \eta \\
    V_\textrm{xxzz} &= -\frac{2}{3} \eta
\end{align}

\subsubsection{Initial boundary value problem}
Introducing the simplifications to the conservation of mass, conservation of momentum and orientation equation results in the system of equations 
\begingroup
\allowdisplaybreaks
\begin{align}
    \label{eq:mass_balance_1d}
    & \frac{\dot{\rho}}{\rho}
      = - \frac{1}{X} \frac{\partial v}{\partial x^*} - \frac{\dot{h}}{h}\\
    \label{eq:momentum_balance_1d}
    &\rho \dot{v}
      = \frac{1}{X} \frac{\partial}{\partial x^*} \left(-p(\rho) + V_\textrm{xxxx} D_\textrm{xx} + V_\textrm{xxzz} D_\textrm{zz}\right)
      -  2 \lambda \frac{v}{h} \left( \frac{v}{v_0}\right)^{m-1} \\
    \label{eq:axx_1d}
    &\dot{A}_\textrm{xx}
     = 2(A_\textrm{xx}-A_\textrm{xxxx}) D_\textrm{xx}\\
    \label{eq:ayy_1d}
    &\dot{A}_\textrm{yy}
     =-2 A_\textrm{yyxx} D_\textrm{xx}\\
    \label{eq:axy_1d}
    &\dot{A}_\textrm{xy}
     =(A_\textrm{xy}-2A_\textrm{xyxx}) D_\textrm{xx},
\end{align}
\endgroup
for the solution variables $\rho$, $v$, $A_\textrm{xx}$, $A_\textrm{yy}$ and $A_\textrm{xy}$.
Initially, the values are $\rho_0=\SI{1480}{\kilogram\per\meter\cubed}$, $v_0=0$, $A_\textrm{xx}=0.5$, $A_\textrm{yy}=0.5$ and $A_\textrm{xy}=0$.
The boundary condition for the momentum equation at $x^*=0$ is $v=0$ at all times.
At $x^*=1$, the boundary condition changes after complete filling and is given as 
\begin{equation}
    0=
    \begin{cases}
        -p(\rho) + V_\textrm{xxxx}D_\textrm{xx} + V_\textrm{xxzz}D_\textrm{zz}, & \text{if}\ X < X_\textrm{max}\\
        v                                                                     , & \text{if}\ X = X_\textrm{max}
    \end{cases}.
\end{equation} 
All other boundaries are no-flux boundaries.

The initial boundary value problem is solved numerically in MATLAB using the \emph{pdepe} solver with 40 discretization points for $x^*$ and a dynamic time step (see~\ref{sec:matlab} for implementation details).

\subsection{Press control model}
\label{sec:press_controller}
The physical press follows a press profile given as $m$ pairs of tool gap and corresponding closing velocity $\left((h_0, \dot{h}_0),(h_1, \dot{h}_1),...,(h_\textrm{m}, \dot{h}_\textrm{m})\right)$. 
Eventually the controller switches to force-control in order to limit stresses on mold and press. 
A virtual press-controller is used to mimic this behavior as boundary condition during the compression molding simulation.
As long as the compression force is below the force at switch-over $F_\textrm{max}$, the profile is linearly interpolated to obtain the current press velocity.
After the switch-over increment $l_0$, a simple PI-controller is employed to determine the current velocity 
\begin{equation}
    \dot{h}_\textrm{l+1} = \dot{h}_\textrm{l} + P_\textrm{p} \epsilon_l + P_\textrm{i} \sum_{i=l_0}^l \frac{\epsilon_i}{2} \Delta t 
\end{equation}
from the normalized error
\begin{equation}
    \epsilon_i = \frac{F_\textrm{max}-F_i}{F_\textrm{max}}\dot{h}_\textrm{m}.
\end{equation} 
The normalization ensures reliable force-control through a wide range of simulation parameters with constant control parameters $P_\textrm{p}=P_\textrm{i}=0.5$.

In the specific case of this press rheometer, the press profile is set to a constant closing speed $\left((h_0=\SI{10}{\milli\meter}, \dot{h}_0=-\SI{1}{\milli\meter\per\second}),(h_1=\SI{0}{\milli\meter}, \dot{h}_1=-\SI{1}{\milli\meter\per\second})\right)$
with a maximum compression force of $F_\textrm{max}=\SI{4400}{\kilo\newton}$.

\section{Results}
\begin{figure}[!ht]
    \centering
    \footnotesize
    \includegraphics[width=0.48\textwidth]{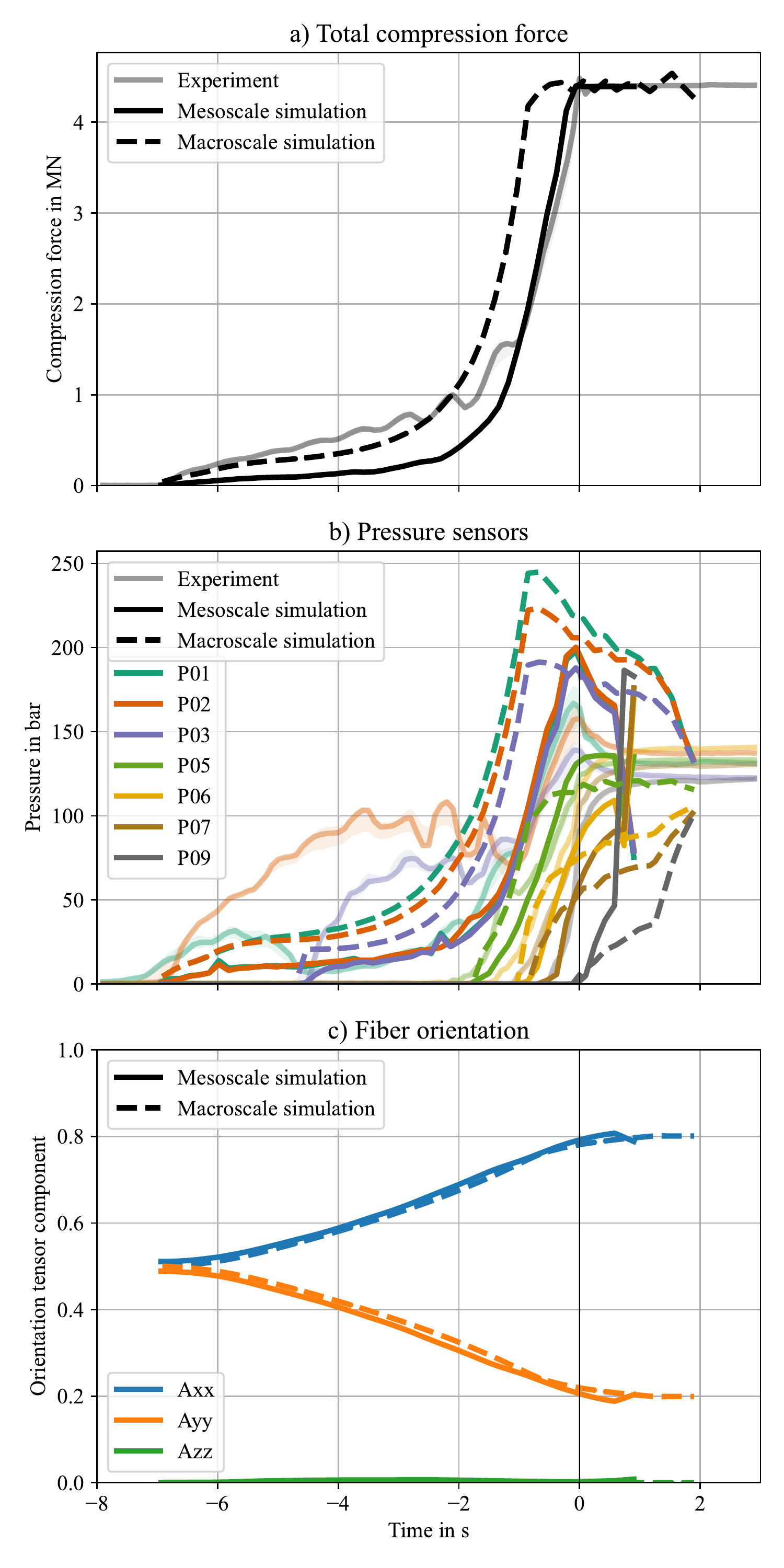}
    \caption{Evolution of process parameters for 25\% initial mold coverage with a shared axis for time, where $t=0$ is the experimental switching point from displacement control to pressure control. The colors and pressure sensor labels in subplot b) refer to the sensor locations indicated in Figure \ref{fig:press_rheometer}.}
    \label{fig:upphgf25}
\end{figure}

Figures~\ref{fig:upphgf75} and~\ref{fig:upphgf25} illustrate the evolution of compression force, pressures and orientation over compression time for press rheometer compression trials and simulations with 75\% and 25\% initial mold coverage, respectively.
The initial stacks are always aligned to the left of the cavity, as depicted in Figure~\ref{fig:press_rheometer}. 
The median experimental results are plotted with light colored solid lines and light colored areas between the first and third quartile, which were obtained from five trials per configuration. 
Additionally, mesoscale simulation results are displayed as solid dark lines and one-dimensional simulation results are displayed as dashed dark lines.
All these results are aligned such that the time $t=0$ describes the switch from displacement controlled mold closing to pressure control. 

\subsection{75\% mold coverage}
\begin{figure}[!ht]
    \centering
    \footnotesize
    \includegraphics[width=0.48\textwidth]{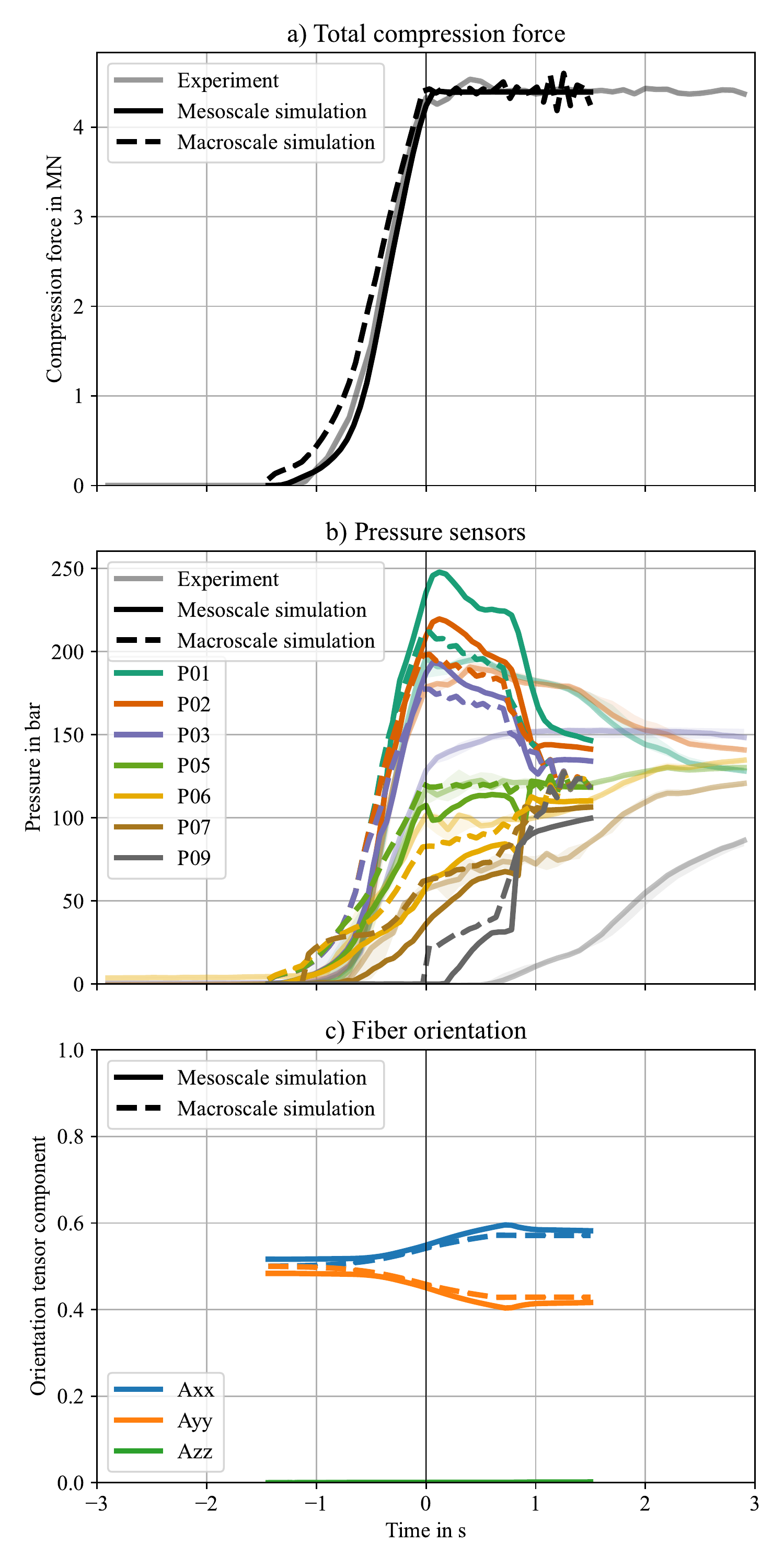}
    \caption{Evolution of process parameters for 75\% initial mold coverage with a shared axis for time, where $t=0$ is the experimental switching point from displacement control to pressure control. The colors and pressure sensor labels in subplot b) refer to the sensor locations indicated in Figure \ref{fig:press_rheometer}.}
    \label{fig:upphgf75}
\end{figure}

An initial stack of 75\% mold coverage is realized by four sheets with dimensions \SI{600}{\milli\meter}~x~\SI{450}{\milli\meter}~x~\SI{4.5}{\milli\meter}.
The experimental compression of these stacks leads to repeatable recordings for the total compression force and pressures, i.e. the area between first and third quartile is narrow (see Figure~\ref{fig:upphgf75} a and b).
The pressure sensor readings (see light solid lines in Figure~\ref{fig:upphgf75} b) are as expected with an increase of pressure from P01 to P09 during the elongational flow. 
However, the sensors do not immediately converge to the nominal pressure $\bar{p} = \SI{122}{\bar}$ after the mold filling is completed at about \SI{1.5}{\second} after reaching maximum compression force.

The mesoscale simulation and the macroscale reference model predict a compression force similar to the experimental recording (see Figure~\ref{fig:upphgf75} a).
The simulated pressure curves feature a bend at $t=0$ due to the transition to a pressure controlled mold closure. 
A second bend occurs when the mold is completely filled and pressure sensors converge towards the nominal pressure $\bar{p}$. 
The spread between P01 and P09 is larger in the mesoscale simulation than in the macroscale model. 
Both simulations overestimate the pressure difference between P01 and P02 slightly and underestimate the total time to complete mold filling. 
A likely cause for both effects is a loss of material through the mold gap at the left side of the tool in the experiments due to the high pressure.

The simulated fiber orientation of the high fidelity mesoscale model gives similar results to the one-dimensional model based on Jeffery's equation (see Figure~\ref{fig:upphgf75} c). 
The mesoscale model has a small initial orientation preference towards the flow direction, which is an artifact of the initial stack generation.
It predicts a reduction of the orientation preference as soon as the mold is completely filled, because bundles at both ends of the mold are forced to a orientation parallel to the wall by non-local effects~\cite{Meyer2020}.
Noticeably, the evaluated $A_{zz}$ component remains negligible, because the \SI{25}{\milli\meter} long fiber bundles are highly constrained between the mold walls, which are located less than \SI{4.5}{\milli\meter} apart.
Therefore, the assumption of planar orientation in the one-dimensional model holds. 
Further, the bundle architecture deforms without significant through-thickness shear, making the perfect plug-flow assumption of the one-dimensional model applicable in this case.

\subsection{25\% mold coverage}
\begin{figure}[!ht]
    \centering
    \footnotesize
    \includegraphics[width=0.9\columnwidth]{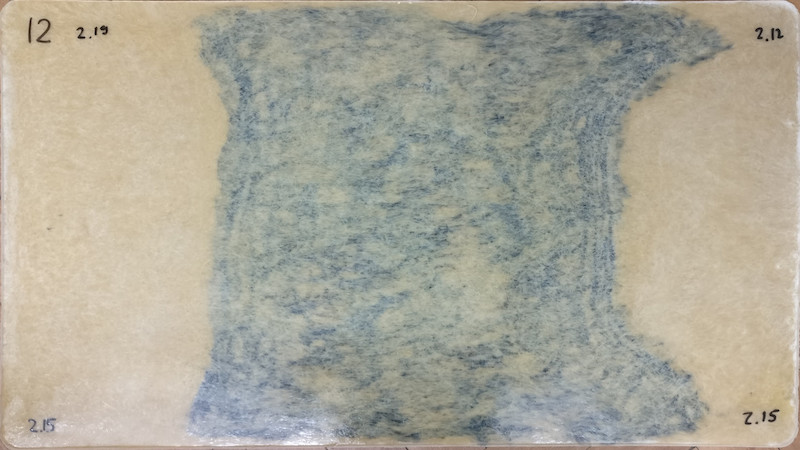}
    \caption{Top view of a molded plaque with a colored sheet at the stack center. 
    The colored sheet has not been stretched over the entire length of the tool (as a plug-flow assumption would suggest), but is transported with relatively low stretch (see Figure \ref{fig:press_rheometer} for initial sheet placement).}
    \label{fig:photo25}
\end{figure}

\begin{figure}[!ht]
    \centering
    \footnotesize
    \includegraphics[width=\columnwidth]{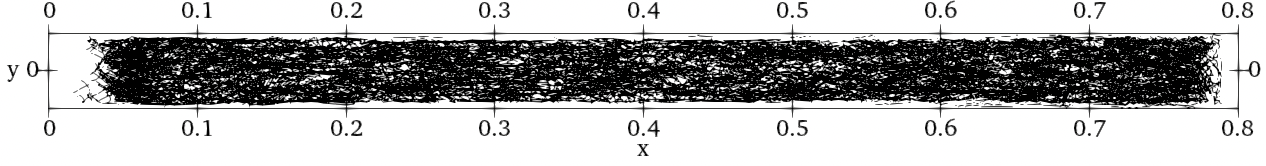}
    \caption{Top view of simulated fiber bundles, which were initially located at the stack center (see Figure \ref{fig:bundles25_detail} for a side view of initial placement). Compared to the experimental results, the simulation predicts significantly more stretch in flow direction.}
    \label{fig:bundles25_final_top}
\end{figure}

\begin{figure}[!ht]
    \centering
    \footnotesize
    \includegraphics[width=0.48\columnwidth]{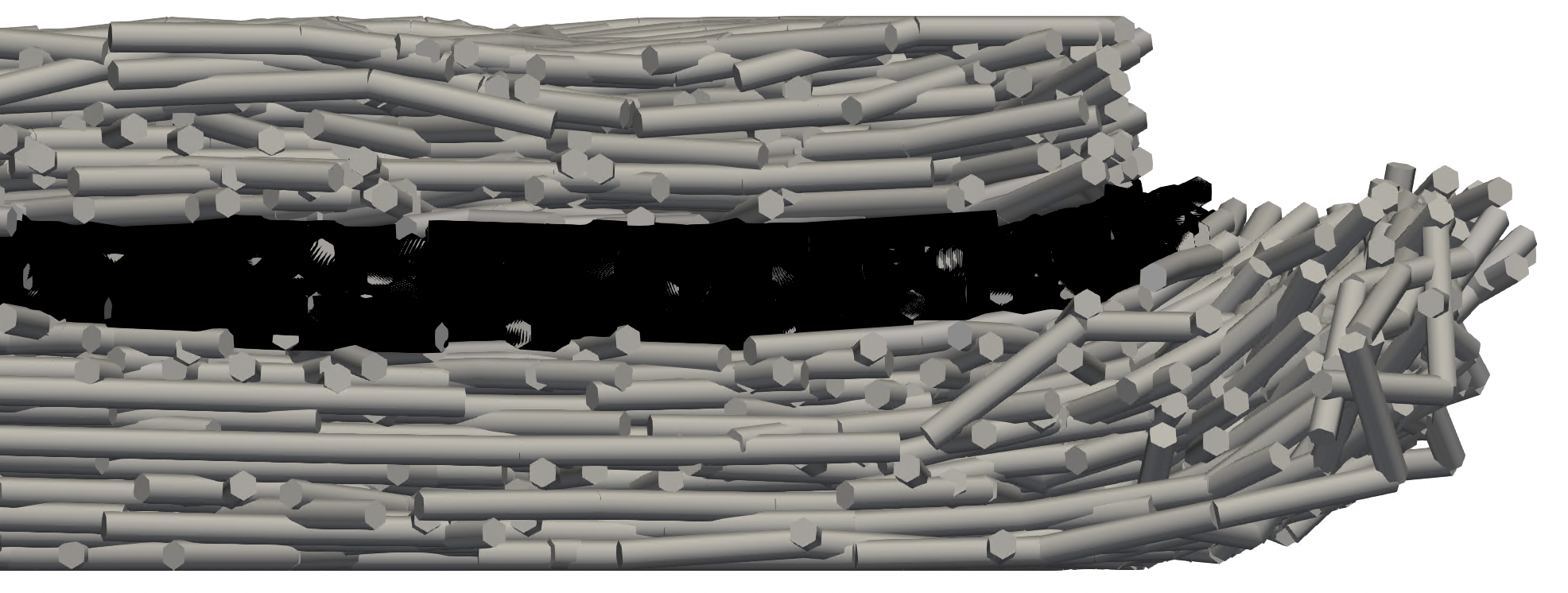}
    \caption{Detail view of the bundle architecture for 25\% mold coverage close to the flow front shortly after the start of the compression process.
    Fiber bundles, that are located initially at the stack center, are colored black.}
    \label{fig:bundles25_detail}
\end{figure}

Compared to the previous case, experimental results feature a larger uncertainty and the ranking of recorded pressure levels is not as expected. 
The leftmost sensor P01 records a pressure drop with pressures that are significantly lower than P02 and even P03.
An initial stack of 25\% mold coverage is realized by nine sheets with dimensions \SI{200}{\milli\meter}~x~\SI{450}{\milli\meter}~x~\SI{8.5}{\milli\meter}. 
This behavior is not correctly reproduced by the simulation models.
The computed forces of the macroscale reference model are too large and result in premature switch to the pressure controlled phase and subsequently in an over-prediction of fill time. 
The mesoscale simulation yields reasonable results at the end of filling, but also fails to predict the severe pressure drop in sensors P01 and P02. 
A cause for the deviation is possibly a deformation mode that is significantly different from ideal plug-flow. 
To investigate this deviation from the plug-flow assumption, plaques with a colored central sheet were manufactured.
A photo of such a molded plaque is given in Figure~\ref{fig:photo25}.
While a perfect plug-flow would result in a homogeneously stretched black layer, the photo shows that the black sheet is shifted and not fully stretched along the flow path.
Areas outside black region feature compressed fiber bundles with distinct flow marks separating those regions. 
Figure~\ref{fig:bundles25_final_top} shows a corresponding top view of predicted positions for fiber bundles, which were initially positioned at the stack center.
The simulation shows a lack of black colored fiber bundles from the central sheet at both ends, albeit to a much lower extend than observed experimentally. 
Even though the mesoscale simulation predicts some shear in the fiber bundle architecture shortly after compression start due to the inhomogeneous temperature (see Figure~\ref{fig:bundles25_detail}), the extent does not agree with experiments. 
However, the similar simulated orientation state between both models highlights once again that Jeffery's basic model agrees well with a direct simulation model for highly confined planar flow.



\section{Discussion}
\label{sec:discussion}
The results show a reasonable agreement between a one-dimensional macroscale model, a mesoscale model with individual resolved fiber bundles, and experimental results for a thin SMC stack with 75\% mold coverage. 
Unlike a mere comparison of compression forces, the comparison of pressures in the mold allows a more detailed evaluation of contributions to the total compression force by mold friction and viscous elongation. 
The observed pressure reduction from P01 to P09 is expected and indicates a correct proportion of those contributions in both models.

For a thick stack with a small mold coverage of 25\%, the results deviate significantly. 
A likely reason for this deviation is that the deformation kinematics differ notably from ideal plug-flow conditions. 
In that case, the one-dimensional reference model cannot predict accurate results, as the entire formulation is based on a plug-flow hypothesis.
The mesoscale simulation shows initial shear contributions due to a non-isothermal temperature distribution in the stack and the initial deformation mode shown in Figure~\ref{fig:bundles25_detail} agrees well with observations from literature~\cite{Odenberger.2004}.
However, the mesoscale model is also not able to fully explain the unexpected pressure histories, shift of the central sheet and formation of areas with compressed fibers at both ends of the mold.

\begin{figure}[!htpb]
    \centering
    \footnotesize
\begingroup%
  \makeatletter%
  \providecommand\color[2][]{%
    \errmessage{(Inkscape) Color is used for the text in Inkscape, but the package 'color.sty' is not loaded}%
    \renewcommand\color[2][]{}%
  }%
  \providecommand\transparent[1]{%
    \errmessage{(Inkscape) Transparency is used (non-zero) for the text in Inkscape, but the package 'transparent.sty' is not loaded}%
    \renewcommand\transparent[1]{}%
  }%
  \providecommand\rotatebox[2]{#2}%
  \newcommand*\fsize{\dimexpr\f@size pt\relax}%
  \newcommand*\lineheight[1]{\fontsize{\fsize}{#1\fsize}\selectfont}%
  \ifx\svgwidth\undefined%
    \setlength{\unitlength}{198.42519685bp}%
    \ifx\svgscale\undefined%
      \relax%
    \else%
      \setlength{\unitlength}{\unitlength * \real{\svgscale}}%
    \fi%
  \else%
    \setlength{\unitlength}{\svgwidth}%
  \fi%
  \global\let\svgwidth\undefined%
  \global\let\svgscale\undefined%
  \makeatother%
  \begin{picture}(1,0.57142857)%
    \lineheight{1}%
    \setlength\tabcolsep{0pt}%
    \put(0,0){\includegraphics[width=\unitlength,page=1]{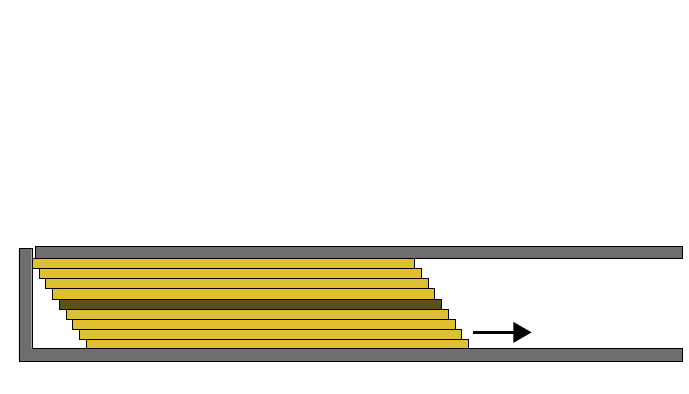}}%
    \put(0.04392454,0.50979909){\color[rgb]{0.36470588,0.31764706,0.10588235}\makebox(0,0)[lt]{\lineheight{1.25}\smash{\begin{tabular}[t]{l}Squeeze mechanism\end{tabular}}}}%
    \put(0,0){\includegraphics[width=\unitlength,page=2]{mechanisms.pdf}}%
    \put(0.04392454,0.23172768){\color[rgb]{0.36470588,0.31764706,0.10588235}\makebox(0,0)[lt]{\lineheight{1.25}\smash{\begin{tabular}[t]{l}Slide mechanism\end{tabular}}}}%
  \end{picture}%
\endgroup%

    \caption{Possible deformation modes.}
    \label{fig:mechanisms}
\end{figure}

Possible deformation modes for the investigated SMC are either a \emph{squeeze mechanism}, in which the central sheet is pushed out or a \emph{slide mechanism}, in which the stack is sheared on the hotter lubricated bottom mold (see Figure~\ref{fig:mechanisms}). 
Both mechanisms would explain the pressure drop of leftmost sensors by an empty region due to the shift of sheets to the right.
The squeeze mechanism would require a transition to a preferred flow of outer hot layers during the compression process, otherwise Figure~\ref{fig:photo25} would show a concentration of the black colored sheet at one end of the mold. 
The slide mechanism is deemed more likely and would explain areas with compressed fiber bundles, as the complete fill of the empty regions requires a compression of the outermost sheets.
Areas of compressed and regular fiber bundles have been previously reported for the same material~\cite{Schemmann.2018}. 

The flow kinematic and resulting pressures could be caused by a yield stress or by a sticking behavior at the mold surface that has to overcome a threshold before entering a hydrodynamic friction regime. 
Further development of characterization methods for the friction model considering temperature changes are required in the future.
Also, the mesoscale simulation model might benefit from distinguishing individual sheets to represent the stack more accurately. 
However, this would require a finer numerical resolution and a method to merge sheets, as they are smeared together in more complex flow scenarios.

Additional sources of error in the one-dimensional macroscale model are the application of Shaqfeh and Fredrickson's~\cite{Shaqfeh.1990} equations, which are originally intended for semi-dilute suspensions only.

Thermal conductivity $\kappa$ is measured only transverse to sheets, even though it is actually an anisotropic tensor with different (orientation dependent) in-plane properties. 
However, the scalar treatment is considered an acceptable simplification, because the heat transfer in this application is dominated by the transverse heat flow and in-plane heat conduction is rather small compared to in-plane convection.

It is remarkable that Jeffery's equation yields results close to a high fidelity direct simulation of individual fiber bundles at mesoscale. 
Jeffery's model has been improved with diffusion terms that drive the orientation towards a more isotropic state, because random fiber collisions lead to a net motion towards more random orientations~\cite{Folgar.1984}.
These diffusion terms were developed with a particular focus on short fiber injection molding.
In SMC compression molding however, a diffusion term of the form $\dot{\bm{A}}_\textrm{D} \propto \dot{\gamma} \left(\bm{I} - 3\bm{A}\right)$ with $\dot{\gamma} = \sqrt{2\bm{D}':\bm{D}'}$ would generate a positive component $\bm{A}_\textrm{zz}$ component.
This component is suppressed for long fibers by the narrow gap between molds, as seen in the direct simulations. 
Hence, Jeffery's basic model seems better suited for planar SMC simulation than its derivations with empirical diffusion terms. 

\section{Conclusions and Outlook}
First, this study reports characterization of key properties for non-isothermal, non-Newtonian compressible behavior of UPPH glass fiber SMC. 
The thermal properties include transverse heat conductivity and conductance from mold to SMC, which were determined from temperature measurements in an SMC stack. 
The viscosity of the paste is measured in a plate-plate rheometer and shows typical power-law behavior.
The hydrodynamic mold friction is computed from pressure differences in an instrumented press rheometer and shows a power-law behavior as well.
The press rheometer is also used to obtain a tabulated expression for the relation between pressure and volumetric compression.

Flow in a press rheometer is then simulated with a high-fidelity mesoscale direct bundle simulation as well as a one-dimensional macroscale reference model utilizing Jeffery's equation for fiber orientation. 
For thin SMC stacks with a large mold coverage of 75\%, both models are able to reproduce compression force and pressure sensor recordings. 
In contrast to a comparison of total compression force only, pressure sensor recordings verify that the contributions from mold friction and anisotropic viscous elongation are in right proportions. 
However, thick stacks of the investigated SMC with small initial mold coverage experience deformations that deviate significantly from ideal plug-flow assumptions.
The one-dimensional reference model cannot describe this deformation by design, but even the detailed mesoscale model is not able to fully predict this behavior.
To predict a sliding mechanism, the mesoscale model may be enhanced by an advanced temperature dependent friction model with a threshold for slipping or a differentiation between sheets.  

For planar SMC flow of thin stacks, Jeffery's equation agrees well with the computed reorientation of the direct mesoscale simulation. 
The original formulation without additional diffusion terms can be recommended for planar, plug-flow dominated SMC molding, because fibers are constrained by the molds and have only marginal out-of-plane orientation components. 

\section{Author Contributions} 
Conceptualization: N.M., A.H., L.K.; 
methodology: N.M., S.I.; 
software: N.M.; 
validation: N.M. and S.I.; 
investigation: N.M. and S.I.; 
resources: N.M., F.H. and L.K.; 
data curation: N.M.; 
writing--original draft preparation: N.M; 
writing--review and editing: N.M., S.I., A.H., F.H., and L.K.;
visualization: N.M.; 
supervision: A.H., F.H. and L.K.;

\section{Acknowledgment}
The research documented in this manuscript has been funded by the Deutsche Forschungsgemeinschaft (DFG, German Research Foundation), project number 255730231, within the International Research Training Group “Integrated engineering of continuous-discontinuous long fiber reinforced polymer structures“ (GRK 2078). 
The support by the German Research Foundation (DFG) is gratefully acknowledged.


\appendix
\section{Tabulated compaction behavior}
\begin{table}[!htbp]
    \centering
    \footnotesize
    \begin{tabular}{lr}
    \toprule
    Hencky strain & Pressure in \si{\mega\pascal} \\
    \midrule
    -0.0000  &     0.0\\
    -0.0770  &     6.3\\
    -0.1098  &    12.6\\
    -0.1325  &    18.9\\
    -0.1496  &    25.3\\
    -0.1638  &    31.6\\
    -0.1749  &    37.9\\
    -0.1840  &    44.2\\
    -0.1923  &    50.5\\
    -0.1982  &    56.8\\
    -0.2029  &    63.2\\
    -0.2073  &    69.5\\
    -0.2116  &    75.8\\
    -0.2167  &    82.1\\
    -0.2219  &    88.4\\
    -0.2270  &    94.7\\
    -0.2317  &   101.1\\
    -0.2349  &   107.4\\
    -0.2378  &   113.7\\
    -0.2407  &   120.0\\
    \bottomrule
    \end{tabular}
    \caption{Relation between Hencky strain and pressure}
    \label{tab:eos}
\end{table}

\section{Implementation details of the mesoscale direct bundle simulation}
The direct bundle simulation described in Section \ref{sec:dbs} and~\citeauthor{Meyer2020}~\cite{Meyer2020} is implemented via several user subroutines in a Coupled Eulerian Lagrangian (CEL) framework in SIMULIA Abaqus Explicit.
A VUEXTERNALDB subroutine manages the overall workflow, i.e. parsing input files before the analysis, building kd-trees for each step, evaluating a stable time step and performing checks. 
A VUFIELD subroutine extracts positions and velocities of nodes in each time step and a VUSDFLD subroutine extrapolates the data to the unique Gaussian point of EC3D8RT and T3D2 elements.
The viscosity of the matrix is computed in a VUVISCOSITY subroutine. 
Finally, a VDLOAD subroutine is used to evaluate equations~\eqref{eq:stokes_generalization}, ~\eqref{eq:relative_velocity}, and~\eqref{eq:bundle_body_force}.
Additionally, a VUINTERACTION subroutine is used to implement Equation~\eqref{eq:tangential_stress} and a VUAMP subroutine realizes the press controller described in Section \ref{sec:press_controller}.

The computation time (wall-clock time) for a full solution is up to \SI{73}{\hour} at 25\% mold coverage (\SI{8}{\second} simulated time) on a workstation with a 16-core Intel Xeon E5-2667 v2 @ \SI{3.3}{\giga\hertz}.

\section{Implementation details of the one-dimensional macroscale reference model}
\label{sec:matlab}
The set of equations~\eqref{eq:mass_balance_1d},~\eqref{eq:momentum_balance_1d},~\eqref{eq:axx_1d},~\eqref{eq:ayy_1d},~\eqref{eq:axy_1d} are reformulated to 
\begin{equation}
    C \left(x^*, t, \vec{s}, \frac{\partial \vec{s}}{\partial x^*} \right) \frac{\partial \vec{s}}{\partial t} = \frac{\partial}{\partial x^*} \vec{f} \left(x^*, t, \vec{s}, \frac{\partial \vec{s}}{\partial x^*} \right) + \vec{s} \left(x^*, t, \vec{s}, \frac{\partial \vec{s}}{\partial x^*} \right)
\end{equation}
for a solution variable vector $\vec{s} = \left(\rho, v, A_\textrm{xx}, A_\textrm{yy}, A_\textrm{xy} \right)^\top$. 
The coupling matrix is defined as 
\begin{equation}
    C = \textrm{diag}\left(\frac{1}{\rho}, \rho, 1, 1, 1 \right), 
\end{equation}
the flux is given by 
\begin{equation}
    \vec{f} = \frac{1}{X}
    \begin{pmatrix}
       -v \\[0.5em]
       -p(\rho) + V_\textrm{xxxx} \frac{1}{X} \frac{\partial v}{\partial x^*} + V_\textrm{xxzz} \frac{\dot{h}}{h} \\[0.5em]
       0 \\[0.5em]
       0 \\[0.5em]
       0
    \end{pmatrix},
\end{equation}
and the source term is given by 
\begin{equation}
\vec{s} = 
    \begin{pmatrix}
       -\frac{\dot{h}}{h} \\[0.5em]
       -2 \lambda \frac{v}{h} \left( \frac{v}{v_0}\right)^{m-1} \\[0.5em]
       2(A_\textrm{xx}-A_\textrm{xxxx}) \frac{1}{X} \frac{\partial v}{\partial x^*} \\[0.5em]
       -2A_\textrm{yyxx}) \frac{1}{X} \frac{\partial v}{\partial x^*} \\[0.5em]
       (A_\textrm{xy}-A_\textrm{xyxx}) \frac{1}{X} \frac{\partial v}{\partial x^*}
     \end{pmatrix}.
\end{equation}

During time integration, the flow front is updated according to 
\begin{equation}
    X^{l+1} = X^{l} + v^l(x^*=1) \Delta t
\end{equation}
with $l$ denoting the current time step index. 
The pressures are evaluates as 
\begin{equation}
    \sigma_\textrm{zz} = -p(\rho) + V_{zzxx} \frac{1}{X} \frac{\partial v}{\partial x^*}  + V_\textrm{zzzz} \frac{\dot{h}}{h}
\end{equation}
and integrated to compute the total compression force 
\begin{equation}
    F = W \int_{x^*=0}^X \sigma_\textrm{zz}(\tilde{x}) \textrm{d}\tilde{x}.
\end{equation}

The computation time (wall-clock time) for a full solution is up to \SI{3.5}{\minute} at 25\% mold coverage (\SI{8}{\second} simulated time) on a desktop computer with a 4-core Intel Core i7-3770 @ \SI{3.4}{\giga\hertz}.

\bibliographystyle{elsarticle-num-names} 
\bibliography{references}

\end{document}